\documentclass{emulateapj}

\usepackage{epsfig, verbatim}
\usepackage{subfigure}
\usepackage{graphicx}
\usepackage{amsmath}
\usepackage{natbib}

\shorttitle{Modeling Gamma-Ray Pulsars using Vacuum Dipole Magnetic
Field} \shortauthors{Bai \& Spitkovsky}

\begin{document}


\title{Uncertainties of Modeling Gamma-Ray Pulsar Light Curves using Vacuum Dipole Magnetic Field}


\author{Xue-Ning Bai \& Anatoly Spitkovsky}
\affil{Department of Astrophysical Sciences, Princeton University,
Princeton, NJ, 08544} \email{xbai@astro.princeton.edu,
anatoly@astro.princeton.edu}




\begin{abstract}
Current models of pulsar gamma-ray emission use the magnetic field of a rotating dipole
in vacuum as a first approximation to the shape of plasma-filled pulsar magnetosphere.
In this paper we revisit the question of gamma-ray light-curve formation in pulsars in
order to ascertain the robustness of the ``two-pole caustic" and ``outer gap" models
based on the vacuum magnetic field. We point out an inconsistency in the literature
on the use of the relativistic aberration formula, where in several works the shape
of the vacuum field was treated as known in the instantaneous corotating frame, rather
than in the laboratory frame. With the corrected formula, we find that the peaks in the
light curves predicted from the two-pole caustic model using the vacuum field are less
sharp. The sharpness of the peaks in the outer gap model is less affected by this change,
but the range of magnetic inclination angles and viewing geometries resulting in
double-peaked light curves is reduced. In a realistic magnetosphere, the modification of
field structure near the light cylinder due to plasma effects may change the shape of the
polar cap and the location of the emission zones. We study the sensitivity of the light
curves to different shapes of the polar cap for static and retarded vacuum dipole fields.
In particular, we consider polar caps traced by the last open field lines and compare them
to circular polar caps. We find that the two-pole caustic model is very sensitive to the
shape of the polar cap, and a circular polar cap can lead to four peaks of emission. The
outer-gap model is less affected by different polar cap shapes, but is subject to big
uncertainties of applying the vacuum field near the light cylinder. We conclude that
deviations from vacuum field can lead to large uncertainties in pulse shapes, and a more
realistic force-free field should be applied to the study of pulsar high energy emission.
\end{abstract}


\section{Introduction}

Pulsars are rotating neutron stars (NS) with very strong magnetic
fields ($B\sim10^9-10^{12}$G). Nearly two thousand pulsars are known,
mostly in the radio band. Compton Gamma Ray Observatory (CGRO)
firmly detected seven gamma-ray pulsars, and another three with less
confidence [see \citet{thom04} for a review]. This number is rapidly
increasing with the start of operations of Fermi Gamma-ray Space
Telescope (e.g., \citealp{fermi08, FermiRelease09a, FermiRelease09b, FermiCatalog}).
Gamma-ray light curves of pulsars are typically double-peaked, generally out of
phase with the radio, and have substantial off-peak emission. Several theoretical
models have been developed to explain the nature of this emission, namely, the
polar-cap model \citep{rs75,hte78,dh82,dh96}, the slot-gap model (SG, or two-pole
caustic, TPC for short; \citealp{as79,arons83,mh03,mh04,dr03,dhr04}), and the
outer-gap (OG) model \citep{chr86a,chr86b,ry95,yadi97,crz00}. In these models,
particles are accelerated in the ``gap" regions, where strong electric fields are
developed due to a deficit of charge. Gamma-ray emission is interpreted as the
curvature and/or inverse Compton radiation from ultra-relativistic particles
accelerated in these gaps. The models differ in the location of the gaps in the
magnetosphere. The polar cap model has narrow beam size and has difficulty in
producing extended light curves, whereas SG/TPC and OG models can reasonably well
reproduce double-peaked, extended light curves. Recent works have shown that OG
model can also reproduce the high-energy spectrum for Crab and Vela pulsars
\citep{hirotani07,tc07,tcc07,tcs08}.

Special relativistic effects, such as the aberration of photon emission direction
and photon travel time delay, are important for calculations of gamma-ray light
curves when the emission zone extends far from the NS surface. In the TPC model,
the emission zone is assumed to be along the last open field lines (LOFLs), extending
from the polar cap to some cut-off radius. In the OG model, the emission zone is
along the open field lines and extends from the null charge surface to the light
cylinder\footnote{Recent version of OG model allows inner boundary of emission zone
to shift toward the NS surface [e.g. \citet{tcs08}].} (LC, $R_{\rm LC}=c/\Omega$). In
both models, the relativistic effects are essential to forming ``caustics" in the sky
map, which appear as sharp peaks in the light curve. The caustics arise when photons
emitted from different regions of the magnetosphere happen to arrive to the observer
at the same time. The presence and appearance of caustics is sensitive to both the
relativistic effects and the geometry of the emission zone.

All existing calculations of pulsar gamma-ray light curves assume that pulsar
magnetospheric geometry can be represented by a vacuum dipole magnetic field.
However, the pulsar magnetosphere is filled with plasma \citep{gj69}, and plasma
currents should modify the field structure. The magnetosphere should then consist of
the open and closed field line regions, separated by thin current sheets. In contrast,
all field lines of the vacuum field, including those that travel beyond the LC, are
formally closed, and no current sheets exist. Numerical solutions of the structure
of plasma-filled magnetosphere are now known in the limit of force-free (FF) MHD for
axisymmetric rotators \citep{ckf99, Gruzinov05, Timokhin06, McKinney06}, and, recently, for
three-dimensional oblique rotators as well \citep{as06, cont09}. The FF field clearly
demonstrates the current sheet structure, and its geometry differs substantially from
the vacuum field near the LC [see \citet{as08} for a review]. Hence, the more realistic
FF field geometry can lead to modifications of gamma-ray light curves [see
\citet{bs09a} for preliminary results]. On the other hand, the vacuum field has been
used over the years to obtain light curves that compare very favorably to the existing
data. It can be argued that if the emission comes from the regions in the magnetosphere
that are not too close to the light cylinder, the field geometry there may be well
approximated by the vacuum field. This raises the question of how reliable and robust
are the vacuum field light curves, and whether perturbations introduced by the presence
of plasma in the magnetosphere would strongly affect the result.

Although our ultimate goal is to study gamma-ray emission using the force-free field,
in this paper we concentrate on the modeling of light curves using vacuum magnetic field
only. The results with the FF field will be presented in the companion paper \citep{bs09b}.
We feel it is necessary to clarify a number of points before moving forward. As we try
to reproduce the sky maps and light curves using vacuum field geometry, we find that
there are ambiguities in the literature on the use of the aberration formula. Since the
aberration effect is crucial to the formation of caustics in any field geometry, in this
paper we clarify the applicability of aberration formulas and compare their influences
on the sky maps and light curves. In order to investigate the potential effects of plasma
on the formation of light curves, we also study the sensitivity of the vacuum light curves
to variations in the shape of the polar cap. Even if the field geometry in the bulk of the
magnetosphere could be approximated by the vacuum field, it is the behavior of the field
lines near the light cylinder that determines the shape of the polar cap and thus the
location of the magnetospheric emission zones. The plasma effects near the light cylinder
can then undermine light curve modeling that uses the polar caps of the vacuum field. In
this paper we show that the appearance of the sky map and light curves is indeed very
sensitive to both the field geometry and the geometry of the emission zones, which
suggests that revisiting theoretical models with the more realistic FF field is essential.

This paper is structured as follows. We begin with the vacuum magnetic field formulas
for pulsar magnetosphere in section 2, and then discuss the effect of aberration in
section 3. In section 4, we construct the shape of the polar caps. We present comparisons
of sky maps and light curves between different aberration formulas as well as different
polar cap shapes for the two-pole caustic and the outer-gap models in section 5. In section
6 we summarize our results.

\section{Vacuum Magnetic Field Formulas}

We will use the vacuum field as an approximation to the magnetic field of the plasma-filled
magnetosphere, where ${\mathbf E}\cdot{\mathbf B}=0$ and the magnetospheric structure is
stationary in the corotating frame (CF). Therefore, in the lab frame (LF), for any specified
magnetic field  ${\bf B}$, there exists an electric field ${\bf E}$ such that:
\begin{equation}
{\bf E}=-\frac{{\mathbf\Omega}\times{\bf r}}{c}\times{\bf B}\
.\label{eq:eb}
\end{equation}
where ${\mathbf\Omega}$ is the angular velocity of the NS, and ${\bf r}$
is the position vector. Equation (\ref{eq:eb}) is of fundamental importance for discussing
the aberration effect (\S3), and is assumed throughout this paper. It does
not apply, however, in the gaps where the high-energy emission is thought to originate.
We address this issue in Appendix B and show that the conclusions of this paper
remain unchanged even in the presence of gaps.

\begin{figure}
    \centering
    \includegraphics[width=75mm,height=85mm]{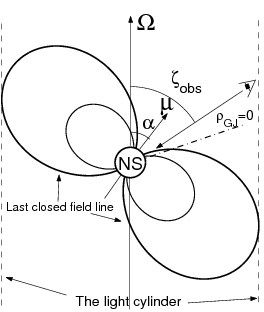}
    \caption{Schematic illustration of magnetospheric geometry. $\alpha$ is the
    inclination angle of the magnetic axis, $\zeta_{obs}$ is the
    observer's viewing angle, and $\rho_{GJ}=0$ marks the surface of zero
    charge density [$\rho_{GJ}\simeq-{\bf{\Omega}}\cdot{\bf{B}}/(2\pi c)$].}\label{fig:1}
\end{figure}

We will consider two commonly used formulas for the vacuum magnetic field, namely, the
static dipole and the retarded dipole. For the static dipole, the field geometry is
assumed to be the same as in a non-rotating dipole field, which is rigidly attached
to the rotating pulsar. The field expression is
\begin{equation}
{\bf B}=\frac{1}{r^3}[3(\vec{\mu}\cdot\hat{r})\hat{r}
-{\mathbf\mu}]\ ,\label{eq:static1}
\end{equation}
where $\vec{\mu}$ is the magnetic dipole moment vector and $\hat{r}$ is the radial unit
vector. For a pulsar rotating along $\hat{z}$ axis with angular velocity $\Omega$ and
magnetic inclination angle $\alpha$, the time evolution of magnetic moment vector is
\begin{equation}
\vec{\mu}(t)=\mu(\sin\alpha\cos\Omega t\hat{x}+\sin \alpha\sin\Omega
t\hat{y}+\cos\alpha\hat{z})\ \label{eq:static2}.
\end{equation}
The static dipole is not a full solution of the field of the rotating dipole in vacuum. As such, 
it can be interpreted as known either in the corotating or the lab frame.
As an approximation, we assume that the magnetic field of the static dipole
is valid in the lab frame. The field geometry is sketched in Fig. \ref{fig:1}.

The full solution of the electromagnetic field of a rotating magnetic dipole is known as
the retarded dipole formula (e.g., \citealt{jack75}):
\begin{equation}
{\bf B}=-\bigg[\frac{\vec{\mu}(t)}{r^3}+
\frac{\dot{\vec{\mu}}(t)}{cr^2}+\frac{\ddot{\vec{\mu}}(t)}{c^2r}\bigg]+{\bf
rr}\cdot\bigg[3\frac{\vec{\mu}(t)}
{r^3}+3\frac{\dot{\vec{\mu}}(t)}{cr^2}+\frac{\ddot{\vec{\mu}}(t)}{c^2r}\bigg]\
.\label{eq:retard}
\end{equation}

When $r$ is small, the retarded dipole field configuration is almost the same as that
of the static dipole. The deviation increases as the radius approaches the LC. One can
find the expressions in Cartesian coordinates in \citet{crz00}. The retarded formula is
valid in the LF, and not in the CF.

Another vacuum formula that can be used to represent the magnetosphere is the Deutsch
field (\citealt{deut55}, \citealt{ml99}), which includes corrections due to the finite
size of the star $R_N$. These corrections are of second order in $R_N/R_{\rm LC}$, and
can be ignored for most pulsar parameters. Therefore, we will not consider the Deutsch
field further\footnote{In addition to the magnetic field, both the retarded dipole and the Deutsch field solutions have the associated electric field. Retarded dipole solution satisfies
${\mathbf E}\cdot{\mathbf B}=0$, while the Deutsch field generally has
${\mathbf E}\cdot{\mathbf B}\neq0$ due to unipolar induction. Since we are using
the vacuum magnetic field to approximate the FF field geometry, we discard these
electric fields and use equation (\ref{eq:eb}) instead.}.

\section{Relativistic Effects in Pulsar Radiation}

Pulsar gamma-ray radiation is believed to originate from
curvature or inverse Compton emission from relativistic particles
(electrons and positrons). In the extremely high electromagnetic
field, these particles travel along magnetic field lines in a
frame where electric field vanishes\footnote{In the force-free 
magnetosphere,  ${\bf E}\cdot {\bf B}=0$
[cf. eq.(\ref{eq:eb})] and $|{\bf E}|<|{\bf B}|$. Therefore, at
any point we can always find an instantaneous frame in which the
electric field locally vanishes. This is not assured, however, outside the
LC for fields in vacuum.}. Emission direction from these highly relativistic
particles, therefore, coincides with their direction of motion. In order to
collect the pulsar emission to obtain a light curve, we need to account for
relativistic effects including the aberration of photon emission direction\footnote{
The aberration of light describes the change in the direction of photon propagation
between different inertial frames. In the context of this paper, it refers to the
correction of the photon emission direction relative to the direction of magnetic field in
the LF.} and the correction of photon travel time (time delay).

The time delay effect is straightforward to calculate. The difference of light
travel time for photons emitted from different regions of the magnetosphere
results in the delay in rotational phase at which the photon is observed
(\citealt{ry95})
\begin{equation}
\phi_d=-{\bf r}\cdot\hat{\eta}/R_{\rm LC}\ ,\label{eq:delay}
\end{equation}
where ${\bf r}$ is the position of the emission point, $\hat{\eta}$ is the
aberrated direction of photon emission (unit vector, in LF). The total phase of
emission is $\phi=-\phi_{em}+\phi_d$, where $\phi_{em}$ is the phase of $\hat{\eta}$.
In order to determine $\hat{\eta}$ from any location in the magnetosphere, we need
to consider the aberration effect. As the direction of particle motion depends on
the direction of the magnetic field, one has to be careful about the frame in which
the field is known. The vacuum field formulas are in general valid in the LF as
discussed in section 2. However, it was commonly implicitly assumed in the literature
that these fields are valid in the instantaneous corotating frame, leading to
discrepancies. For clarity, we will distinguish the following three frames:

1. The lab frame (LF), which is the inertial observer's frame in which the pulsar is
rotating around the $\hat{z}$ axis.

2. The corotating frame (hereafter CF), which is a non-inertial frame that corotates
with the pulsar and is the frame in which the field pattern is steady. It is related
to the LF by the coordinate transformation
\begin{equation}
\begin{split}
x'&=x\cos\Omega t+y\sin\Omega t\ ,\\
y'&=-x\sin\Omega t+y\cos\Omega t\ ,\\
z'&=z,\qquad t'=t\ ,\\
\end{split}\label{eq:cflf}
\end{equation}
where the prime denotes the coordinate in the CF.

3. The instantaneous corotating frame (hereafter ICF), which is a {\it local} frame of
an inertial observer instantaneously moving at the corotation velocity. Therefore,
it is defined only inside the light cylinder.

We will discuss the aberration effect in these three frames separately. We pose the
question as follows. The pulsar magnetic field configuration is known in the LF at
$t=0$ and the field pattern corotates with the star. The electric field in the LF
obeys equation (\ref{eq:eb}). The photon emission direction at any position is along
the magnetic field in a frame where the electric field vanishes. Our goal is to
calculate the photon emission direction seen in the LF. We will do the calculation
in three different frames and show that we reach the same result, as expected. From now
on, we add superscripts ``$C$" and ``$I$" to denote fields in CF and ICF respectively.
Fields with no superscripts will always refer to fields in the LF.

First, we do the LF calculation. Consider the motion of an emitting
particle in the pulsar magnetosphere. The force-free condition requires
${\bf E}+\vec{\beta}_0\times{\bf B}=0$, where $\vec{\beta}_0$ is the
normalized velocity of the particle relative to $c$, or the direction of
the emitted photon in the LF. This equation, when combined with equation
(\ref{eq:eb}), implies
\begin{equation}
\vec{\beta}_0=f{\bf B}+\vec{\beta}_{\rm rot}\ ,\label{eq:lfabr}
\end{equation}
where $\vec{\beta}_{\rm rot}={\bf\Omega}\times{\bf r}/c$ is the normalized
corotation velocity, and $f$ is a coefficient. For the emitting particle, we
have $|\vec{\beta_0}|\rightarrow1$, from which $f$ can be determined. Solving
equation (\ref{eq:lfabr}) with $|\vec{\beta_0}|=1$ fixes the emission direction
$\hat\eta=\vec{\beta}_0$. We note that there are always two solutions associated
with equation (\ref{eq:lfabr}), one associated with particles traveling along
the magnetic field line in the CF, the other with particles traveling in the
opposite direction\footnote{At the LC, $f=0$ is an obvious solution where the particle
corotates at the LC. This solution corresponds to the ``backward moving" solution.
However, we always pick the other solution, where the particle moves outward.}.

Second, we consider the calculation in the CF. The relation of the electromagnetic
field between the CF and the LF is \citep{schiff39, gron84}
\begin{equation}
{\bf B}^C={\bf B}\ ,\qquad {\bf E}^C={\bf E}+\vec{\beta}_{\rm rot}\times{\bf
B}\ .\label{eq:ebcflf}
\end{equation}
According to equation (\ref{eq:eb}), we have ${\bf E}^C=0$. The photon emission in
this frame is thus along ${\bf B}^C={\bf B}$. We emphasize here that CF and LF are
related by a coordinate transformation rather than Lorentz transformation. In Appendix A,
we show that the aberration formula calculated from this frame has the same expression
as equation (\ref{eq:lfabr}).

Finally, consider aberration in the ICF. The magnetic field in LF and ICF are related by
a Lorentz transformation. Using equation (\ref{eq:eb}) for electric field in the LF, we
obtain
\begin{equation}
B_t^{I}=B_t\ ,\quad {\bf B}_p^{I}={\bf B}_p/\gamma\ ,\quad {\bf
E}^{I}=0\ ,\label{eq:ebicflf}
\end{equation}
where $\gamma\equiv(1-\beta_{\rm rot}^2)^{-1/2}$ is the Lorentz factor of corotation,
$B_t$ and ${\bf B}_p$ denote the toroidal and poloidal components of the magnetic field.
Note that in ICF the poloidal magnetic field is smaller. Since ${\bf E}^I=0$, the photon
emission direction is thus
\begin{equation}
\eta^{I}_t=\pm B_t/B'_0\ ,\qquad \vec{\eta}^{I}_p=\pm {\bf
B}_p/\gamma B'_0\ ,\label{eq:emicf}
\end{equation}
where $B'_0\equiv\sqrt{B_t^2+(1-\beta_{\rm rot}^2)B_p^2}$ is the total magnetic field
strength in the ICF, and plus/minus sign corresponds to emission along/opposite to the
magnetic field line. To get back to the LF, one should perform a Lorentz transformation
to aberrate the direction of the photon $\hat{\eta}^{I}\rightarrow\hat{\eta}$; the
formula is \citep{dr03}
\begin{equation}
\hat{\eta}=\frac{\hat{\eta}^{I}+[\gamma+(\gamma-1)(\vec{\beta}_{\rm rot}\cdot\hat{\eta}^{I})
/\beta_{\rm rot}^2]\vec{\beta}_{\rm rot}}{\gamma(1+\vec{\beta}_{\rm rot}\cdot\hat{\eta}^{I})}\
.\label{eq:abr}
\end{equation}
Substituting equation (\ref{eq:emicf}) into equation (\ref{eq:abr}) we find
\begin{equation}
\eta_t=\frac{\beta_{\rm rot} B_p^2\pm B_tB'_0}{B_0^2}\
,\qquad\vec{\eta}_p=\frac{\pm B'_0-\beta_{\rm rot} B_t}{B_0^2}{\bf B}_p\
,\label{eq:icfabr}
\end{equation}
where $B_0\equiv\sqrt{B_t^2+B_p^2}$ is the total magnetic field strength in the LF.

We note that although the ICF is defined only within the LC, equation (\ref{eq:icfabr}) is
valid anywhere. In fact, the ICF is only one special instantaneous frame in which
${\bf E}^{I}$ vanishes. We can construct other instantaneous frames where the electric field
is zero, and equation (\ref{eq:icfabr}) is a general result. For example, we can also choose
the ${\bf E}\times{\bf B}$ instantaneous drift frame (IDF). One can show that by transforming
the LF field to IDF to calculate the aberration leads to the same result as equation
(\ref{eq:icfabr}). Simple algebra shows that equations (\ref{eq:icfabr}) and (\ref{eq:lfabr})
are exactly equivalent as expected. Therefore, our proof is complete.

In many of the previous studies (e.g., \citealp{ry95,yadi97,crz00,dr03,dhr04}), it was
implicitly assumed that the vacuum field, given by any formula in section 2, is valid
in the ICF instead of LF. The following equation was used instead of equation (\ref{eq:emicf}):
\begin{equation}
\eta^I_t=\pm B_t/B_0\ ,\qquad \vec{\eta}^I_p=\pm {\bf B}_p/B_0\
.\label{eq:wrabr1}
\end{equation}
The corresponding photon emission direction is then
\begin{equation}
\eta_t^\times=\frac{\beta_{\rm rot} B_0\pm B_t}{B_0\pm\beta_{\rm rot} B_t}\ ,\qquad
\vec{\eta}_p^\times=\frac{{\bf B}_p}{\gamma(B_0\pm\beta_{\rm rot} B_t)}\
.\label{eq:wrabr2}
\end{equation}

In the limit $\beta\rightarrow0$, equations (\ref{eq:icfabr}) and (\ref{eq:wrabr2}) agree.
The differences between them are of the order ${\cal O}(\beta_{\rm rot}^2)$ when
$\beta_{\rm rot}\ll1$. Since the difference between the two aberration formulas
scales as $\beta_{\rm rot}^2$, then to first order in $r/R_{\rm LC}$ treating the vacuum dipole
field as known in ICF instead of LF, as was done in \citet{dhr04}, is a reasonable
approximation. However, as we shall see in \S5, this introduces significant differences
in the calculated light curves. In the other limit, $\beta_{\rm rot}\rightarrow1$, the two
treatments of aberration result in different photon emission directions. Interestingly, the
backward solution of equation (\ref{eq:icfabr}) agrees with the forward solution of
(\ref{eq:wrabr2}), where the emission direction is perpendicular to the rotation axis.

Since the vacuum field only approximates the FF field, one could choose to treat it as known
in the LF, CF or ICF, as a matter of approximation. However, as shown in the next section, the
last open field lines must be traced either in the LF or in the CF, but not in the ICF. This
makes the treatment in a number of previous works not self-consistent. We will show in \S
5 that the two approaches can lead to substantial differences in the sky maps and thus the
light curves.

\section{Last Open Field Lines and the Shape of the Polar Cap}

Determination of the last open field lines (LOFLs) that separate the open and closed regions
in the magnetosphere is needed to find the shape of the polar cap and the location of the
emission zones. LOFLs are typically found by tracing field lines and checking whether they
close inside or outside the LC. This tracing must be done either in LF or in CF, but not in the
ICF. From equation (\ref{eq:ebicflf}), the toroidal field in ICF coincides with LF or CF field,
but the poloidal component is $\gamma$ times smaller. ICF field is thus ill-defined at the LC.
As $R\rightarrow R_{LC}$, ${\bf B}^I_p\rightarrow0$, the field is purely toroidal, and no field
line actually crosses the LC. Also, since ICF is not a global frame, local tracing of the field
does not guarantee the tracing of a consistent set of field lines as seen from other frames due
to aberration. Thus, the shape of the polar cap in ICF or in any other instantaneous frame is not
well defined\footnote{Another way to think about field lines is that these are trajectories of
force-free particles in the magnetosphere. As seen in the CF, the electric field is zero, and the
trajectory just traces the field line. In the LF, eq. (\ref{eq:lfabr}) determines the trajectory,
including the rotation of the field pattern and motion along the field line. Since
${\bf B}^C={\bf B}$, these two viewpoints are consistent with each other. The trajectory in the ICF,
besides being ill-defined at the LC, does not, in general, trace the same field line as in the LF
or CF. Additional discussion of particle trajectories can be found in Appendix B of \citet{bs09b}.}.

\begin{figure}
    \centering
    \includegraphics[width=93mm,height=70mm]{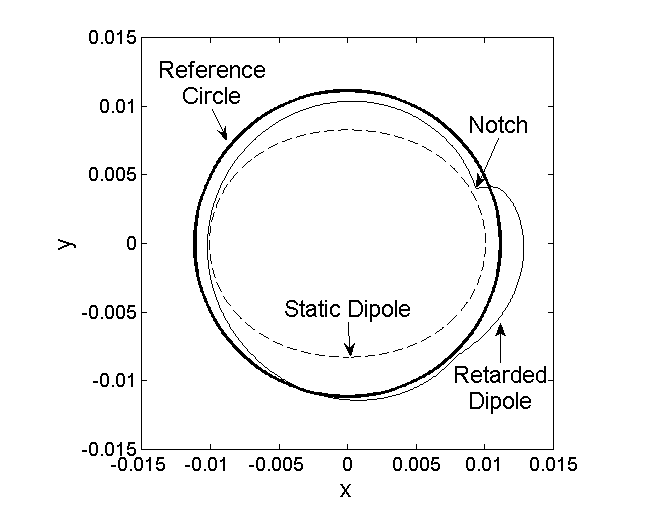}
    \caption{Polar cap shapes for the static (dashed) and retarded (thin solid)
    dipole fields. The thick solid line is a reference circle corresponding to
    $\theta_0=\arcsin{\sqrt{R_N/R_{\rm LC}}}$.
    }\label{fig:2}
\end{figure}

To find LOFLs, we use 3rd order Runge-Kutta integration to trace the magnetic field lines in
the LF. Bisection method \citep{dhr04} is used to find the magnetic colatitude
$\theta_m^{\rm{rim}}$ of the rim of the polar cap for every magnetic azimuth $\phi_m$. We show
the resulting polar cap shape for both the static and retarded dipole fields in Fig. \ref{fig:2}
for magnetic inclination angle $\alpha=60^\circ$. The polar cap shape of the retarded dipole as
seen in Fig. \ref{fig:2} is commonly found in the literature [e.g., \citealp{yadi97,crz00,dhr04}].
Therefore, in these works the retarded dipole field was traced in the LF, but for the computation of
aberration the field was treated in the ICF, leading to an inconsistency.

The vacuum field geometry appears similar to the FF field geometry near the NS surface,
but the two are substantially different near the LC \citep{as06}. The polar cap shape, as determined by LOFLs, is
thus sensitive to the field structure near the LC. This may cause significant uncertainties in the
geometry of the emission zones. As we show in the companion paper \citep{bs09b}, the polar cap shape
in the FF field is, in fact, more circular and also larger than in the vacuum field. Thus, a circular
polar cap may better characterize the emission zone geometry when vacuum field is used. \citet{dr03}
considered a static dipole field with a circular polar cap, given by
$\theta_m=\theta_0 =\arcsin{\sqrt{R_{NS}/R_{\rm LC}}}$. In this paper, we also consider such a
circular polar cap which is indicated as a reference circle in Fig. \ref{fig:2}.  We will refer to
the polar cap found by tracing LOFLs as ``traced polar cap," to distinguish it from the circular
polar cap.

In order to parameterize different field lines, we define open volume coordinates on the polar cap.
The magnetic colatitude of the polar cap rim $\theta_m^{{\rm{rim}}}(\phi_m)$ is generically a
function of azimuth. For any point on the NS surface at the magnetic colatitude $\theta_m$ and
azimuth $\phi_m$, we define the open volume coordinate of this point to be ($r_{\rm ov},\phi_m$),
where $r_{\rm ov}\equiv\theta_m/\theta_m ^{{\rm{rim}}}(\phi_m)$. Therefore, the rim of the polar
cap and LOFLs correspond to $r_{\rm ov}=1$. All open field lines have $r_{\rm ov}<1$ while all closed
lines have $r_{\rm ov}>1$. We emphasize that open volume coordinates should be defined for magnetic
fields in the LF or CF\footnote{The polar cap shape of a retarded dipole has a ``notch" (Fig. \ref{fig:2}). A special
treatment of open volume coordinates in this region was given in \citet{dhr04}. We do not
use this method, as the differences in the sky maps due to better resolution of the notch region are
small.}.

\section{Sky Maps and Light Curves from Vacuum Dipole Field}

In this section we construct the sky maps and calculate the associated
light curves for pulsar gamma-ray emission in vacuum fields. We consider both the
two-pole caustic (TPC) and the outer-gap (OG) models for static and
retarded dipole fields. In each case, we compare the results using different
aberration formulas [i.e., equations (\ref{eq:icfabr}) and (\ref{eq:wrabr2})].
We also compare the results using different polar cap shapes.

The TPC model is an extended version of the slot-gap (SG) model,
where emission has until recently been assumed to come from a thin sheet centered
on the LOFLs, i.e., $r_{\rm ov}^0=1$.\footnote{In the extended version of
the TPC model, the emission zone is localized between $r_{\rm ov}=1$
and $r_{\rm ov}=1-\delta$, where $\delta$ describes the thickness of the gap
\citep{vhg09}. We do not use this definition in order to directly compare
with earlier works (e.g., \citealp{dr03,dhr04}).} The emission zone extends
from above the polar cap along the LOFLs up to a certain cut-off radius,
where emissivity is assumed to drop to zero. The cut-off is described by
$r_{\rm max}$, the distance to the center of the NS, and $R_{\rm
max}$, the cylindrical radius to the rotational axis. These two
parameters constrain the extent of the emission zone.
For the OG model, the emission zone is assumed to come from a layer
in the open field line region beyond the null charge surface (NCS)
where Goldreich-Julian charge density equals to zero\footnote{Recent
development of OG model allows the inner boundary of the emission
zone to be shifted inside the NCS (e.g., \citealp{tcs08}). For our
purpose, to compare different aberration formulas and polar cap shapes,
it suffices to consider emission beyond the NCS only.} [$\rho_{\rm
GJ}\simeq{\bf\Omega} \cdot{\bf B}/(2\pi c)$]. We adopt $r_{\rm
ov}^0=0.9$ for the center of this layer. The emission can extend to
$R_{\rm max} \sim R_{\rm LC}$, and no extra cut-off is needed. In
all the cases, emissivity is assumed to be constant along the field
lines in the emission zone. For different field lines, emissivity is
weighed by a Gaussian function centered at $r_{\rm ov}^0$, with
width $\sigma=0.025$ (\citealt{dhr04}).

The radiation from the emission zones, when projected to the observer
as the NS rotates, produces the light curve. It is convenient to collect
all photons from the emission zones as they fall on the sky and plot the
intensity as a function of stellar phase $\phi$ and the observer's viewing
angle $\xi_{\rm{obs}}$. The light curve is then obtained by a cut trough
this sky map at the observer's viewing angle $\xi_{\rm{obs}}$.

Mathematically, the sky map is a map from the emission zone to the
sky coordinates. At fixed $r_{\rm ov}$, the emission zone is a two
dimensional manifold, which can be parameterized by ($\phi_m, l$),
where $l$ is the length of the magnetic field line starting from the
polar cap region, and $\partial/\partial{l}$ is along the direction
of the magnetic field in units of $R_{LC}$. With constant emissivity along field lines,
the Jacobian determinant of this transformation determines the intensity on the
sky map:

\begin{equation}
I(\phi, \xi_{\rm obs})\propto\bigg|\det{\frac{\partial{(\phi,
\xi_{\rm obs})}} {\partial{(\phi_m, l)}}}\bigg|\
.\label{eq:intensity}
\end{equation}
When the inverse of the Jacobian matrix is singular, infinite magnification is
reached, and $I\rightarrow\infty$. This means that light rays emitted from the
neighborhood around the position $(\phi_m, l)$ arrive to the observer at the same
time, which greatly strengthens the intensity. This is the analog of caustics in
optics and strong gravitational lensing. In the case of the sky map, a ``caustic"
is often understood as regions on the sky map with strong enhancement, where the
determinant in ({\ref{eq:intensity}}) is much greater than 1. The caustics on the sky map then correspond
to peaks in the light curve. There is no guarantee of the existence of caustics on
the sky map mathematically, and a general field structure or a variation in emission
region geometry does not necessarily produce caustics. In the following subsections we
will show examples of how the properties of caustics change as we use different
field geometries, shapes of the polar caps and aberration formulas.

\subsection{Two-pole caustic model with static dipole field}

The TPC model with a static dipole field was considered by
\citet{dr03}. They used a circular polar cap, with
$\theta_m=\arcsin{\sqrt{R_N/R_{LC}}}$, rather than a traced polar
cap. In the calculation of the sky map, they traced the field lines
using the vacuum dipole formula (\ref{eq:static1}). Therefore, the
field was treated as in the LF, but equation (\ref{eq:wrabr2})
was used for aberration, making the treatment not self-consistent.
In this subsection we reproduce this result and compare with other
sky maps obtained from self-consistent treatment using the static
dipole field.

\begin{figure}
    \centering
      \includegraphics[width=90mm]{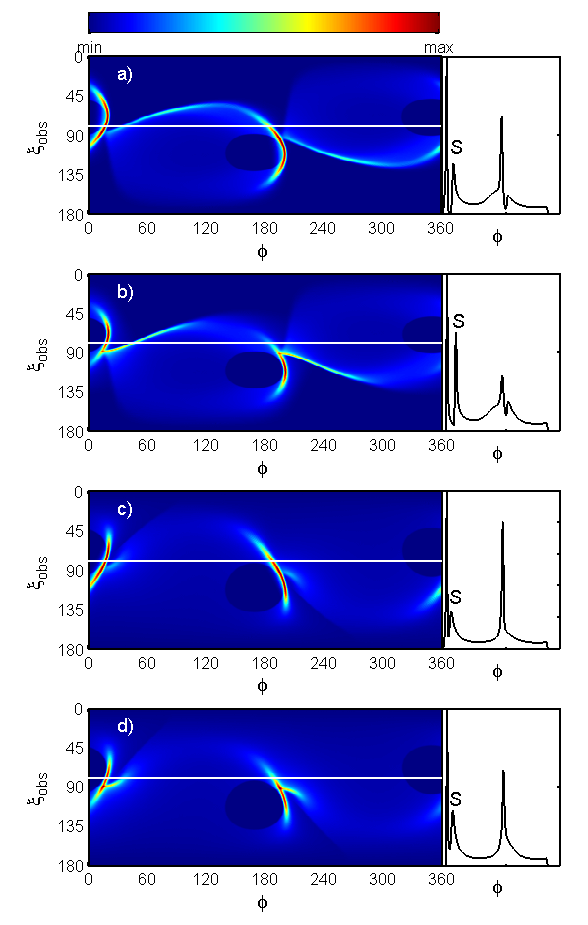}
  \caption{Sky maps (left) for the TPC model using static dipole field with
  inclination angle $\alpha=70^\circ$. (a),(b) use traced polar cap while (c)
  (d) use circular polar cap. Aberration effect is treated consistently in (b)
  and (d), while in (a) and (c), equation (\ref{eq:wrabr2}) is used for aberration.
  Panel (c) is a reproduction of \citet{dr03}. In all panels, lines are traced to
  $r_{\rm{max}}=0.90$. Dark circles indicate the polar cap. Color scale of each
  panel is independent, set by the minimum and maximum counts on the sky map.
  To the right of each panel we plot the light curves for observer's viewing angle
  $\xi_{\rm obs}=80^\circ$.}\label{fig:3}
\end{figure}

In Fig. \ref{fig:3}, we show four sky maps and light curves using
the static dipole field with inclination angle $\alpha=70^{\circ}$.
In (a) and (b) we use the traced polar cap, while in (c) and (d) we
use the circular polar cap. Aberration formula (\ref{eq:wrabr2}) is
used in panels (a) and (c), which is not self-consistent. Other two
panels have self-consistent aberration [using formula (\ref{eq:icfabr})].
The case considered by \citet{dr03} corresponds to Fig. \ref{fig:3}c.
Representative light curves are plotted on the right of each panel in
Fig. \ref{fig:3}; the observer's viewing angle is chosen to be $80^\circ$.

In all panels of Fig. \ref{fig:3}, bright arc-like regions are
caustics, which translate to sharp peaks in the light curve. As far
as the bulk appearance of the sky maps is concerned, Figs.
\ref{fig:3}a and \ref{fig:3}b look similar, while Fig. \ref{fig:3}c
and \ref{fig:3}d look similar. On the other hand, in each of the
corresponding pairs, the details of the caustics differ. Therefore,
the overall appearance of the sky map is very sensitive to the shape
of the polar cap, while the choice of aberration formula is important for
determining the detailed features of the light curves. To some extent,
the inconsistent treatment of magnetic field in the aberration formula in
a number of earlier works may not be critical, but at the current level
of very precise light curve measurement with the Fermi telescope, detailed
differences in the light curves can be distinguishable.
We now discuss the sky maps and the corresponding light curves for the four cases.

We begin with Fig. \ref{fig:3}c, which is a reproduction of Fig. 4
in \citet{dr03}. In the sky map, we have two main caustics near the
polar caps, which are very prominent and sharp. Their
$\xi_{\rm{obs}}$ extends from $50^\circ$ to $110^\circ$ in the sky
map. Right behind the main peak, a small hump in the sky map leads
to a ``subdominant" peak (labeled with ``$S$" in the light curves).
As a result, two sharp peaks and a small, subdominant peak right
behind the first peak is present in the light curve. This light
curve was used to explain Vela's gamma-ray light curve by
\citet{dr03}.

While Fig. \ref{fig:3}d looks similar to Fig. \ref{fig:3}c, a change
in the aberration formula causes some quantitative differences. The
``subdominant" peak becomes stronger, and the extent of the main caustics
becomes smaller. As a result, it is less probable to have a double-peaked
light curve and more difficult to reproduce Vela's profile.

Figs. \ref{fig:3}a,b, where LOFLs were traced to get polar cap shape, look
quite different from Figs. \ref{fig:3}c,d, which used a circular polar cap.
Compared to Fig. \ref{fig:3}c, the shape of the main caustic is more
curved, and the original subdominant peaks become strong and extended caustics.
Also, the emission zones resulting in two peaks now cover a smaller fraction
of the sky map. Now it is almost impossible to reproduce Vela's light curve with this inclination angle.

In sum, the double-peak feature for gamma-ray pulsars is hard to reproduce
with the static dipole field, with the exception of case (c), where the
inconsistent treatment of magnetic field in the aberration formula is used.
This situation is true within a wide range of interesting inclination angles
$\alpha$ (adjusting $r_{\rm{max}}$ and $R_{\rm{max}}$, or changing
$r_{\rm ov}^0$ to smaller value does not help either). This example
demonstrates how the sky maps and the resulting light curves
are affected by the aberration formula and the shape of the polar caps.
Meanwhile, the differences between the panels (b) and (d) in Fig.
\ref{fig:3} characterize the uncertainties of sky maps and light curves
calculated using the static dipole field.

\subsection{Two-pole caustic model with retarded dipole field} {\label{s:tpcret}}

\begin{figure}
    \centering
      \includegraphics[width=90mm]{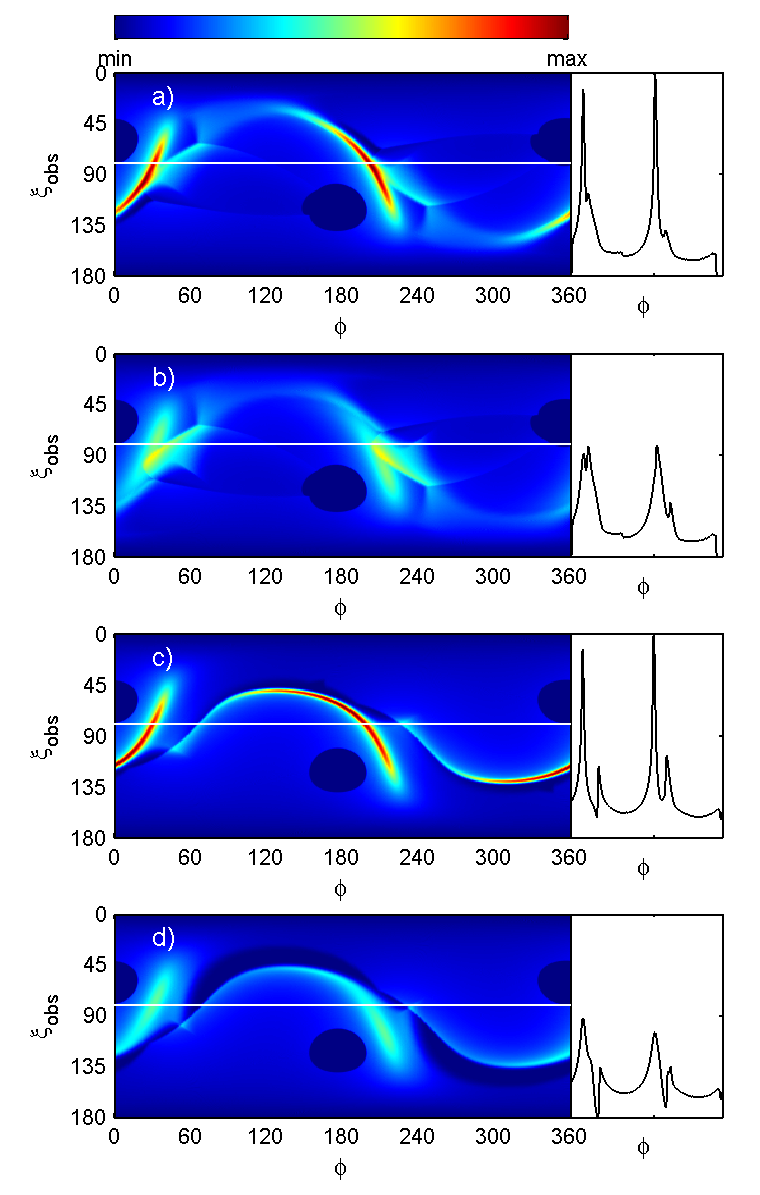}
  \caption{Sky maps (left) for the TPC model using the retarded dipole field with
  inclination angle $\alpha=60^\circ$. (a),(b) use traced polar cap while (c),(d)
  use circular polar cap. Aberration effect is treated consistently in (b) and (d),
  while in (a) and (c), equation (\ref{eq:wrabr2}) is used for aberration. Panel
  (a) is a reproduction of \citet{dhr04}. Lines are traced to
  $r_{\rm{max}}=1$ and $R_{\rm{max}}=0.75$ in (a), (b) and (d); to
  $r_{\rm{max}}=1.2$ and $R_{\rm{max}}=0.8$ for (c). Dark circles indicate
  the polar cap. Color scales of (a) and (b) are the same to demonstrate the
  weakening of the caustic caused by aberration treatment. Similar for (c),
  (d). On the right of each panel shows the light curves at observer's viewing
  angle $\xi_{\rm obs}=80^{\circ}$.}\label{fig:5}
\end{figure}

Next, we consider the two-pole caustic model with a retarded dipole
field (e.g., \citealp{dhr04}). We choose the inclination angle
$\alpha=60^{\circ}$ and show the sky maps and light curves in Fig.
\ref{fig:5}. As in the last subsection, we also consider four
cases. In all of them, the retarded dipole formula is considered valid
in the LF, and is also traced in the LF. In (a) and (b), we use the
traced polar cap, and the circular polar cap is used for the other two.
In (a) and (c), the aberration formula assumes that the field is
in the ICF (which is not self-consistent), while for the other two
figures, the corrected aberration formula [eq. (\ref{eq:lfabr})] is
used. We have chosen $\xi_{\rm obs}=80^{\circ}$ for all light curves in
Fig. \ref{fig:5}.

Fig. \ref{fig:5}a is a reproduction of the sky map and light curve
of \citet{dhr04}. There are two strong
caustics formed near the phase of two poles that extend over a wide
range of observer's viewing angles. Just behind the first peak
there is a small hump, caused by the overlap of emission from two
poles. A small step-like drop in the middle corresponds to a weak
discontinuity in polar cap rim \citep{dhr04}.

Fig. \ref{fig:5}b is the corrected version of Fig. \ref{fig:5}a,
where aberration is treated self-consistently. Remarkably, the two
strong caustics become blurred into weak and wide enhancements
at the same phases. Since the caustics are weak, the brightest
region in the sky map is mainly caused by the overlap of
emission from both poles. In the light curve, we thus have two wide
peaks, and a large fraction of radiation is from the off-peak
phases. We have also explored the sky maps with other inclination
angles, and find that the strong caustics appear in fewer combinations
of inclination angles and viewing geometries (see Appendix \ref{appatlas}).
The weak and wide caustic structure is generic for this TPC model using
the retarded dipole field, and it has difficulty in producing sharp peaks
in the light curve\footnote{If one changes $r_{\rm ov}^0$ to some smaller
value (e.g., $r_{\rm ov}^0=0.9$), the caustics become less affected
by the aberration formula. A full sky map from $r_{\rm ov}^0=0.9$
using the retarded dipole field can be found in Fig. 12 of \citet{bs09b}.
The same trend can be found in the OG sky maps shown in \S5.3.
Although choosing smaller $r_{\rm ov}^0$ improves the light curves from
the TPC model, it still suffers from large uncertainties as discussed further in \S{\ref{s:tpcret}}.}.

Fig. \ref{fig:5}c and \ref{fig:5}d have circular polar caps. The
appearance of the sky map is quite different from Fig. \ref{fig:5}a
and \ref{fig:5}b. The caustics are more extended, and it is
possible to get two peaks from the same pole (in total up to
four peaks). Meanwhile, we find the same trend: the inconsistent
treatment of B field in the aberration leads to strong and narrow
caustics; when aberration is corrected, the caustics become
wide and weak.

We see that the shape of the polar cap largely determines the
overall appearance of the sky map, while the details of the caustics
are sensitive to the aberration formula. The geometry of the emission
zone is mainly controlled by the shape of the polar cap. This example
confirms that the shape of the polar cap (or the emission zone
geometry) is one major source of uncertainty in the current modeling
of gamma-ray pulsar light curves using TPC model (e.g., \citealp{dhr04}).

\begin{figure}
    \centering
      \includegraphics[width=90mm]{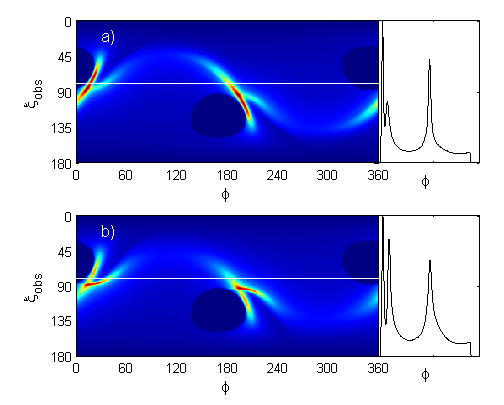}
  \caption{Sky maps (left) for the TPC model using the static dipole field with
  inclination angle $\alpha=60^\circ$ for comparison with Fig. \ref{fig:5}.
  Circular polar cap is used. Aberration effect is treated consistently in (b),
  while in (a), equation (\ref{eq:wrabr2}) is used for aberration. Lines are
  traced to $r_{\rm{max}}=0.9$. Dark circles indicate the polar cap. On the right
  of each panel shows the light curves at observer's viewing angle
  $\xi_{\rm obs}=80^{\circ}$.}\label{fig:6}
\end{figure}

We can also directly compare the effect of different field geometries
by using the same circular polar cap shape. Since we have previously chosen
different inclination angles in Fig. \ref{fig:3} (static dipole,
$\alpha=70^\circ$) and Fig. \ref{fig:5} (retarded dipole, $\alpha=60^\circ$)
in order to reproduce previous works by \citet{dr03,dhr04}, in Figure \ref{fig:6} we plot the sky map for the static dipole field with
$\alpha=60^\circ$ and a circular polar cap to facilitate the comparison with Fig. \ref{fig:5}.
We consider both treatments of aberration in Fig. \ref{fig:6}. One can
compare Fig. \ref{fig:5}c, \ref{fig:5}d with Fig. \ref{fig:6}a, \ref{fig:6}b.
The only difference between the two pairs of the sky maps is different field
geometry. Since the retarded and static dipole fields are similar near
the star, one might expect that the resulting sky maps would be similar as
well. However, we see substantial differences in the appearance of
the sky maps and the light curves, and the differences are already very
prominent in regions that are not far from the star, as seen from the structure of
the caustics. In particular, the main caustics form closer to the star and
appear stronger in the case of the static dipole field than in the
retarded dipole. Even though the caustics form at low altitude, they are very
sensitive to the choice of field structure because caustics are a chance
overlap of emission from different regions. This analysis shows
that the appearance of the sky map is very sensitive to the field geometry
itself, and the deviation of the vacuum field geometry from the more
realistic FF field should be another major source of uncertainty in the
modeling of gamma-ray pulsar light curves.

\subsection{Outer gap model with retarded dipole field}

The OG model with a retarded dipole field was studied by many
authors (e.g., \citealp{ry95,yadi97,crz00,dhr04,tc07,tcc07, tcs08}).
As before, we consider four cases in Fig. 5: panels (a) and (b)
with a traced polar cap, and c) and d) with a circular polar cap. Consistent
treatment of aberration is done in (b) and (d) only.
Case (a) was explored by \citet{ry95,yadi97,crz00,dhr04},
and case (b) was recently explored by \citet{tc07,tcc07,tcs08}, who
have corrected the aberration formula. These studies were able to
reproduce some of the gamma-ray pulsar light curves.
Here we focus on the comparison between the four cases to address the
uncertainties of the OG model.

\begin{figure}
    \centering
      \includegraphics[width=90mm]{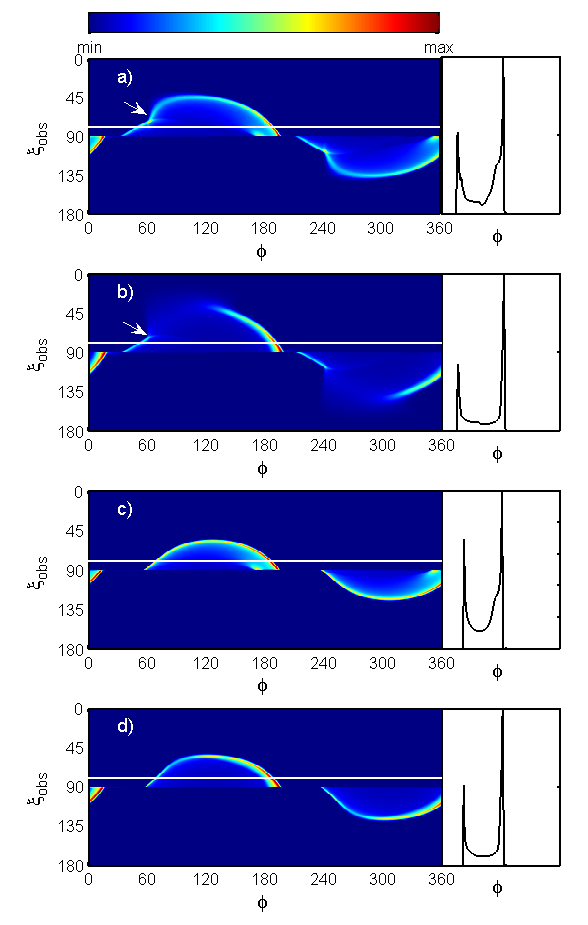}
  \caption{Sky maps (left) of the outer gap model using the retarded dipole
  field at inclination angle $\alpha=65^\circ$. (a),(b) use traced polar cap
  while (c),(d) use circular polar cap. Aberration effect is treated consistently
  in (b) and (d), while in (a) and (c), equation (\ref{eq:wrabr2}) is used for
  aberration. Panel (a) is the same as a number of previous works.
  Lines are traced from the null surface to $r_{\rm{max}}=1.6$ and
  $R_{\rm{max}}=1$ for (a)-(d). The emission zone is centered at
  ${r_{\rm ov}}=0.9$. The arrows in (a) and (b) indicate the effect of
  the ``notch" on the polar cap. On the right of each panel is the light curves
  at observer's viewing angle $\xi_{\rm obs}=80^{\circ}$.}\label{fig:7}
\end{figure}

Fig.~\ref{fig:7} shows the sky maps for the four cases with
inclination angle $\alpha=65^\circ$. We further show the light
curves at $\xi_{\rm{obs}}=80^\circ$ for all four cases to the
right of the sky maps. Our Fig. \ref{fig:7}a is a reproduction of
\citet{ry95,yadi97,crz00,dhr04}. The two peaks are caused by emission
from the same pole. The offset from phase zero and the double peak
profile are clearly present. The arrow indicates the effect of the
``notch" in the polar cap, and the emission from both sides of the notch
contributes to the caustic. In between the double peaks is the
``bridge" emission with weaker intensity. In the plot with corrected
aberration (Fig. \ref{fig:7}b), the regions on the sky map responsible
for the first caustic become less extended. The emission from field lines
on one side of the notch no longer forms caustics. Also, the regions
responsible for the ``bridge" become weaker. Fig. \ref{fig:7}c and
\ref{fig:7}d are the sky maps for the OG model with circular polar
caps. Although the sky maps look different from cases (a) and (b),
the light curves look very similar to their counterparts [(c)
resembles (a), and (d) resembles (b)].

We note that when we correct the aberration formula for the OG model,
the sharpness of the two peaks is less affected than in the TPC
model. This appears odd since the emission zone of OG is close to 
the LC, and one may expect larger difference between the two treatments
of aberration. 
As one can see from Eq.~(\ref{eq:intensity}), the intensity of
the caustics is a differential effect. It depends on the
separation (on the sky map) between the projection of two neighboring
points in the emission zone. Therefore, it is possible that caustic
structures are similar for the two treatments of aberration in
the OG model.

Compared with the TPC model, the appearance of the sky maps and
lights curves in the OG model is less sensitive to the shape of the
polar cap. This is not surprising, because the open field lines not
too close to the LOFL are not sensitive to small displacements in the
polar cap. Therefore, the sky map does not change much when we
replace the traced polar cap with a circular one. This might suggest
that the OG model using the vacuum field is more robust than the
SG model. However, the magnetic field used by the OG model is needed
close to the LC, which itself is still very uncertain. As we have
discussed in the last paragraph of \S5.2, the uncertainties in
the field geometry itself can also cause big differences in the sky
maps. Therefore, the accuracy of the OG model using vacuum field is
still in question.

\section[]{Conclusions}

All current models of pulsar gamma-ray light curves assume that
the pulsar magnetosphere can be represented by the vacuum magnetic
dipole field. The use of the vacuum field should be considered as an
approximation to the field which includes the effects of plasma in the
magnetosphere. In this paper we considered the application of vacuum
field to theoretical models [the two-pole caustic (TPC) and the outer-gap
(OG) models]. Our results show that there are large uncertainties in using
the vacuum field for predictions of gamma-ray emission.

Calculations of high-energy pulsar light curves involve the construction
of a map from the emission zone in the magnetosphere to the sky map. The
appearance of the sky map is thus very sensitive to: 1) the geometry of the
magnetic field; 2) the geometry of the emission zone. On the one hand, the
vacuum field deviates from the FF field, which introduces one major uncertainty
in the direction of emission. On the other hand, the shape of the polar cap is
traced by the last open field lines (LOFLs), and sensitively depends on the field
structure near the LC \citep{dh04}. This causes another large uncertainty
because the geometry of the emission zone is largely determined by the shape of
the polar cap.

Relativistic effects including aberration and time delay are crucial
to the formation of caustics in the sky map. We provide the treatment of the
aberration effect in three different reference frames, namely, the
lab frame (LF), the corotating frame (CF) and the instantaneous
corotating frame (ICF), and show that they can be reconciled. To be consistent,
however, the tracing of the field lines to find the polar cap must be done in
the LF or the CF, but not in the ICF, because the latter is not a global frame.

We compare the sky maps and light curves for the TPC model and the
OG model using different aberration formulas. We find that the appearance
of the caustics in the sky maps is sensitive to the treatment of B field
in the aberration formula. For the TPC model using a retarded dipole field,
we find that instead of having two strong caustics in the sky map,
the corrected aberration formula weakens the caustics, leaving two
wide and weak humps in the sky map. As a result, the conventional TPC
model with retarded dipole field has difficulty in producing sharp peaks
in the light curve. For the OG model using a retarded dipole field and
the corrected aberration formula, the caustic responsible for the
first peak in the light curve is less extended, and there is weaker
``bridge" emission between the two peaks. Recent developments in the
OG model with the corrected aberration formula are still able to produce
reasonable light curves \citep{tc07,tcc07}.

We study the uncertainties in the models of pulsar gamma-ray light curves
by: 1) comparing the sky maps and light curves using different shapes of
the polar cap, namely, the polar cap obtained by tracing LOFLs and
the circular polar cap; 2) comparing the sky maps and light curves using
the same polar cap shape (circular), but with different magnetic
field configurations (static vs.~retarded dipole). We choose
circular polar cap as another possibility because the polar cap of the FF magnetosphere is
more circular \citep{bs09b}. Our results
show that the overall appearance of the sky map is very sensitive to
both factors. We find that for the TPC model, whose emission zone is
centered on the last open field lines, the sky maps and light curves
depend sensitively on the changes in the shape of the polar cap. Up
to four peaks can be present when circular polar cap is used. This
suggests that the reproduction of the Vela's light curve in
\citet{dr03,dhr04} is not robust. For the OG model, whose emission
zone is in the open field line region, we find that the sky maps
and the light curves are not very sensitive to changes in the polar
cap shapes. However, the vacuum field near the LC is unreliable, and
the predictions from the outer gap model are still questionable.
A detailed atlas of gamma-ray light curves for two models is presented
in Appendix C.

In all, we conclude that it is essential to revisit the existing
theoretical models using a more realistic magnetospheric
structure, i.e, the force-free field from numerical simulations
\citep{as06}. In the companion paper \citet{bs09b}, we will present
the sky maps and light curves calculated using the force-free field.

\acknowledgments

We thank Yury Lyubarsky and Jonathan Arons for help and advice. We
also thank our referee for helpful suggestions. This work is supported
by NASA grants NNX08AW57G and NNX09AT95G. AS acknowledges support from
Alfred P. Sloan Foundation Fellowship. XNB acknowledges support from
NASA Earth and Space Science Fellowship.

\appendix

\section[A]{A. Aberration formula in the CF}

We can find the inverse transformation of equation (\ref{eq:cflf})
and write it in the differential form as
\begin{equation}
\begin{pmatrix}dx\\dy\\dz\\dt\end{pmatrix}=
\begin{pmatrix}
\cos\Omega t' & -\sin\Omega t' & 0 & -\Omega y \\
\sin\Omega t' & \cos\Omega t' & 0 & \Omega x \\
0 & 0 & 1 & 0 \\0 & 0 & 0 & 1
\end{pmatrix}
\begin{pmatrix}dx'\\dy'\\dz'\\dt'\end{pmatrix}\ .
\end{equation}

It suffices to consider $t=t'=0$, in which case we also have $x=x'$,
$y=y'$. Dividing each side by $cdt=cdt'$, we obtain the photon
velocity vector
\begin{equation}
\eta_x=\eta'_x-\Omega y/c\ ,\qquad \eta_y=\eta'_y+\Omega x/c\
,\qquad\eta_z=\eta'_z\ .\label{eq:cfabr1}
\end{equation}
In this equation $\vec{\eta}$ is a unit vector that denotes the
photon direction in the LF, while $\vec{\eta}'$ denotes the photon
motion in the CF, but it is NOT a unit vector. This is because
the metric in the CF is not the Minkowski metric, and has
non-diagonal space-time components \citep{schiff39}.

In the case of pulsar gamma-ray emission, $\vec{\eta}'$ is along the
direction of ${\bf B}^C={\bf B}$. Therefore, equation
(\ref{eq:cfabr1}) can be rewritten as
\begin{equation}
\vec{\eta}=f{\bf B}+\Omega\times{\bf r}/c\ .
\end{equation}
This is exactly the same as equation (\ref{eq:lfabr}).

\section[B]{B. Aberration formula in the presence of gaps} \label{appgaps}

In this appendix we relax the FF assumption to allow the presence of gaps in
the magnetosphere. In the gaps, $E_\parallel\neq0$, and equation (\ref{eq:eb})
no longer holds. However, we show that under the following two assumptions,
the aberration formula discussed in this paper still holds in the gaps:
\begin{description}
\item[a)] The magnetospheric structure is stationary in the CF;
\item[b)] Particles move along magnetic field lines in the CF.
\end{description}
Assumption 1) allows us to work in the CF using equation (\ref{eq:cflf}).
Assumption 2) implies that in the CF, the perpendicular electric field must
be zero (while $E_{\parallel}^{CF}$ can be non-zero in the gaps). Transforming
the field back to the LF by applying equation (\ref{eq:cflf}), we obtain
\begin{equation}
\begin{split}
{\mathbf E}_{\perp}&=-\frac{{\mathbf \Omega}\times{\mathbf r}}{c}\times{\mathbf B}\ ,\\
{\mathbf E}_{\parallel}&={\mathbf E}_{\parallel}^{CF}\ .\\
\end{split}\label{eq:abrgap}
\end{equation}
Physically, assumption b) means the gap in the magnetosphere also corotates with
the star, because the ${\mathbf E}\times{\mathbf B}$ drift velocity has a corotation
component everywhere.

Under the above two assumptions,  even if particles are accelerated due to $E_{\parallel}$
in the gaps, their direction of motion, and, hence, the direction of photons they emit, is
always along the B field in the CF. Therefore, the treatment of aberration in this
case is exactly the same as in the Appendix A, and the formula (\ref{eq:icfabr}) is valid
even in the presence of gaps. This analysis can be reproduced in the LF as well.

The above discussions also apply if the magnetosphere rotates differentially with
respect to the NS (e.g., \citealp{Timokhin07a,Timokhin07b}), as long as one replaces
$\Omega$ with the value of the angular velocity in the magnetosphere.

\section[C]{C. Light Curve Atlas} \label{appatlas}
\begin{figure}
    \centering
      \includegraphics[width=150mm]{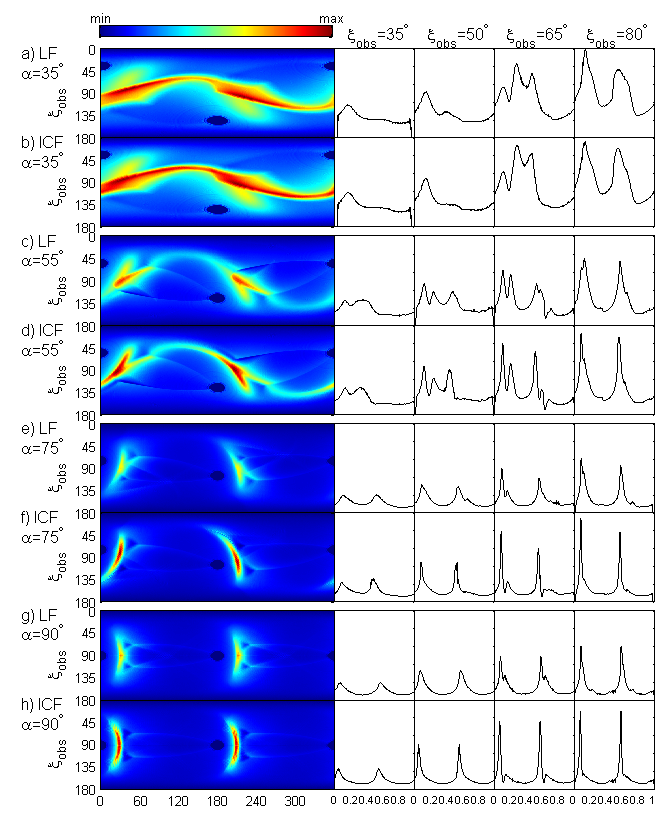}
  \caption{Atlas of sky maps and light curves for the two-pole caustic (TPC) model for
  representative pulsar inclination angles (constant for each row) and for different
  observer viewing angles (arranged in columns). Panels a), c), e), g) show results from
  treating the retarded field in the lab frame, while b), d), f), h) treat the field in
  the instantaneous corotating frame.}
\label{fig:b1}
\end{figure}

\begin{figure}
    \centering
      \includegraphics[width=150mm]{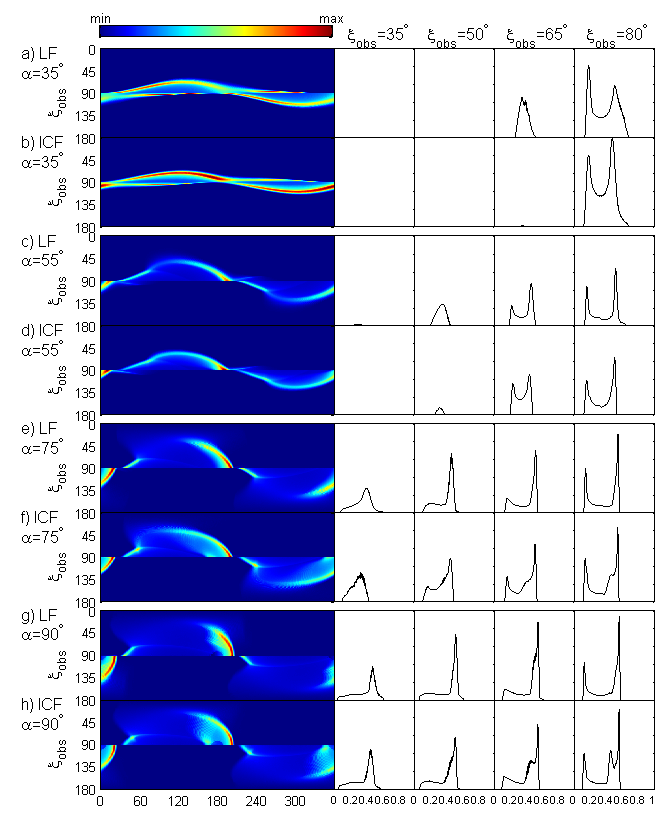}
  \caption{
  Atlas of sky maps and light curves for the outer gap (OG) model for representative
  pulsar inclination angles (constant for each row) and for different observer viewing
  angles (arranged in columns). Panels a), c), e), g) show results from treating the
  retarded field in the lab frame, while b), d), f), h) treat the field in the
  instantaneous corotating frame.}
 \label{fig:b2}
\end{figure}

Knowing how the light curves appear for different viewing angles and magnetic
obliquities can help place constraints on the pulsar geometry. Recently,
\cite{wat09} compiled an atlas of TPC and OG light curves for a range of pulsar
parameters using retarded vacuum field. The field was effectively assumed to be in the
instantaneous corotating frame for the purposes of aberration calculation.
Here we provide a similar atlas, emphasizing the differences brought by the
self-consistent treatment of aberration. In Figs. \ref{fig:b1} and
\ref{fig:b2} we show the sky maps and representative light curves for TPC
and OG models, respectively. For every pulsar inclination angle (horizontal rows)
we show the results for field taken in the lab frame (label LF, subpanels a, c, e, g)
and in the instantaneous corotating frame (label ICF, subpanels b, d, f, h). As was
shown above, only the LF treatment for the retarded vacuum field is self-consistent.
The polar cap is obtained by tracing LOFLs in all cases. The light curve plots have
identical vertical scales at the same obliquity for easy cross-comparison.

The atlas displays the same general trends as seen before: self-consistent aberration
reduces the strength of the caustics in the TPC model, and decreases the amplitude of
the first peak for many observer angles in the OG model.

\end{document}